\documentclass{emulateapj}

\newcommand{\be}{\begin{equation} }
\newcommand{\ee}{\end{equation} }
\newcommand{\bea}{\begin{eqnarray} }
\newcommand{\eea}{\end{eqnarray} }

\slugcomment{Accepted to ApJL: 25 Aug 2008}

\shortauthors{Maness {\it et~al.\/}}

\shorttitle{Millimeter Imaging of HD 32297}

\begin{document}

\title{CARMA Millimeter-Wave Aperture Synthesis Imaging of the HD
32297 Debris Disk}

\author{ 
H. L. Maness\altaffilmark{1},
M. P. Fitzgerald\altaffilmark{2}, 
R. Paladini\altaffilmark{3},
P. Kalas\altaffilmark{1},
G. Duchene \altaffilmark{1,4}, 
James R. Graham\altaffilmark{1}
} 

\altaffiltext{1}{Astronomy Dept, UC Berkeley, Berkeley, CA 94720}
\altaffiltext{2}{Michelson Fellow, IGPP, L-413, Lawrence Livermore
National Laboratory, 7000 East Ave, Livermore, CA 94550}
\altaffiltext{3}{Spitzer Science Center, California Institute of
Technology, Pasadena, CA 91125}
\altaffiltext{4}{Laboratoire d'Astrophysique de Grenoble, CNRS/UJF UMR
5571, F-38041 Grenoble cedex 9, France}

\begin{abstract}
We present the first detection and mapping of the HD 32297 debris disk
at 1.3 mm with the Combined Array for Research in Millimeter-wave Astronomy (CARMA).
With a sub-arcsecond beam, this detection represents the highest
angular resolution (sub)mm debris disk observation made to date. Our
model fits to the spectral energy distribution from the CARMA flux and
new Spitzer MIPS photometry support the earlier suggestion that at
least two, possibly three, distinct grain populations are traced by
the current data.  The observed millimeter map shows an asymmetry
between the northeast and southwest disk lobes, suggesting large
grains may be trapped in resonance with an unseen exoplanet.
Alternatively, the observed morphology could result from the recent
breakup of a massive planetesimal.  A similar-scale asymmetry is also
observed in scattered light but not in the mid-infrared.  This
contrast between asymmetry at short and long wavelengths and symmetry
at intermediate wavelengths is in qualitative agreement with
predictions of resonant debris disk models.  With resolved
observations in several bands spanning over three decades in
wavelength, HD 32297 provides a unique testbed for theories of grain
and planetary dynamics, and could potentially provide strong
multi-wavelength evidence for an exoplanetary system.
\end{abstract}


\section{Introduction}

Debris disks provide the principal means of studying the formation and
evolution of planetary systems on timescales of 10$-$100 Myr.
Evidence for exoplanets in these systems can be found by matching
density variations in debris disks to theoretical models of the
gravitational perturbations caused by planets (e.g., Reche et
al. 2008).  A modest sample of debris disks have now been imaged in
the visible, and many show substructure such as clumps, warps, and
offsets, consistent with dynamical perturbations by massive planets.
However, a wide variety of other mechanisms can produce similar
structures (Moro-Martin et al. 2007, and references therein).  As
different wavebands are sensitive to different grain sizes, which are
in turn subject to different dynamical influences, multi-wavelength
observations offer the most promising path towards definitively
classifying the physical mechanisms at work in these systems.  At
present, only a few debris disks have resolved observations spanning
more than a decade in wavelength.  A particularly critical, though
technologically challenging, deficit of observations lies at
(sub)millimeter wavelengths, which trace large grains primarily
affected by gravitational forces.  To date, bolometer arrays have
resolved 8 debris disks in the (sub)mm (e.g., Holland et al. 1998,
Greaves et al. 1998). However, such low-resolution
($\theta_{\mathrm{beam}} \gtrsim 10''$) single-dish measurements are
limited to the largest, nearest disks.  Higher resolution
interferometric observations are needed to access the larger debris
disk population already imaged at shorter wavelengths. Some pioneering
work has been done in this area; OVRO and PdBI have detected and
resolved two debris disks (Vega: Koerner et al. 2001, Wilner et
al. 2002; HD 107146: Carpenter et al. 2005).  Recently,
\citet{Corder08} mapped HD 107146 with the Combined Array for Research
in Millimeter-wave Astronomy (CARMA) at 1.3 mm, providing the highest
fidelity interferometric debris disk map to date.  Here, we report the
near-simultaneous CARMA detection of HD 32297, the third debris disk
mapped with a (sub)mm interferometer.  With a sub-arcsecond beam, this
detection is the highest angular resolution (sub)mm debris disk
observation made to date.

HD 32297 is a $\sim$30 Myr A-star at $112_{-12}^{+15}$ pc
\citep{Perryman97}, first discovered to host a resolved debris disk
with HST/NICMOS near-infrared (NIR) imaging \citep{Schneider05}.  The
discovery image showed an edge-on debris disk extending to 400 AU
(3.3$''$), with an inner-disk brightness asymmetry inward of 60 AU
($0.5''$).  \citet{Kalas05} subsequently imaged HD 32297 in the
optical, revealing an asymmetric, extended outer disk ($\sim$1700 AU,
15$''$) likely interacting with the interstellar medium.  Later,
\citet{Redfield07} detected circumstellar gas in this system,
reporting the strongest Na I absorption measured toward any known
debris disk. Most recently, \citet{Fitzgerald07}, hereafter F07, and
\citet{Moerchen07} resolved HD 32297 in mid-infrared (MIR), thermal
emission.  Detailed analysis of the spectral energy distribution (SED)
by F07 showed that multiple grain populations may be present in the
disk.  The lack of long wavelength data needed to characterize the
large grain properties of HD 32297 motivated the CARMA observations
presented here.

\section{Observations and Data Reduction}

We observed HD 32297 with CARMA, located 7200 ft above sea level
outside Big Pine, California and consisting of six 10.4-m and nine
6.1-m antennas, previously comprising the OVRO and BIMA arrays. We
used CARMA's D configuration (baselines: 11$-$148 m) on 2008 March 06
and the more extended C configuration (baselines: 26$-$370 m) on 2007
October 26 and 2007 November 08.  We tuned the receivers to a central
frequency of 227 GHz.  The total continuum bandwidth is 1.5 GHz,
contained in three 500 MHz bands.  Our total observation time was 19
hours in good weather with rms path errors of $\lesssim 175 \mu$m and
zenith opacities of $\tau_{230} \lesssim 0.35$.  Throughout our
observations, we used optical offset guiding to maximize pointing
accuracy (Corder, Carpenter \& Wright 2008, in prep).

We calibrated the data using the MIRIAD software package.  We
performed passband calibration with 15 minutes integrations on a bright
quasar (3C 84, 3C 111) observed at the start of each track.  We
derived time-dependent phase solutions from 3 minutes integrations on
J0530+135, observed following each 15 minutes integration on source.  In
addition, we used 1 minutes integrations on a secondary calibrator,
3C 120 (observed every cycle), to test the astrometric
accuracy of the interferometer and the integrity of our phase
solutions.  Finally, we flux-calibrated the data using a 5 minutes
integration on a planet (Uranus, Mars) observed once per track.  The
systematic uncertainty in CARMA's absolute flux scale is $\sim$20\%
(W. Kwon, private communication).

The 1.3 mm map of HD 32297 combining all observations is shown in
Figure \ref{map}.  For this map, we adopted natural weighting to
provide optimal sensitivity and processed the visibility data as a
mosaic to accomodate the heterogeneous array.  We deconvolved the
dirty map using the Steer CLEAN algorithm for mosaics \citep{Steer84},
set to bring the rms in the cleaned region of the residual map to that
measured in an off-source region.  To emphasize possible resolved
structure, we restored the map with a circular beam with radius equal
to that of the semimajor axis of the naturally weighted beam
($\theta_{\mathrm{beam}}=0.9''$).

In addition to the new CARMA data presented here, we also extracted
Spitzer MIPS photometry from director's discretionary time program 225
(PI: G. Schneider).  Using the MOPEX package, we manually calibrated
the 24 $\mu$m data to remove the strong background gradient and
``jailbars'' evident in the pipeline processed image.  The 70 $\mu$m
data did not suffer severe artifacts; in this case, we simply
mosaicked the data. The 160 $\mu$m data are contaminated by a spectral
leak, which occurs for bright sources and is typically corrected using
observations of a calibration star.  However, no such calibration
observation was made in this program; we, therefore, include the 160
$\mu$m data in this analysis only as an upper-limit.  The MIPS fluxes
as derived from aperture photometry are $F_{\nu}$(23.68 $\mu$m) = 0.21
$\pm$ 0.01 Jy, $F_{\nu}$(71.42 $\mu$m) = 0.85 $\pm$ 0.06 Jy,
$F_{\nu}$(155.9 $\mu$m) $<$ 0.46 $\pm$ 0.06 Jy.  In the SED modeling
(\S 4), we adopt calibration uncertainties listed in the MIPS data
handbook and apply a color correction appropriate to the modeled grain
temepratures.

\section{Results}

\begin{figure}
\centering
\includegraphics[width=.45\textwidth,angle=0]{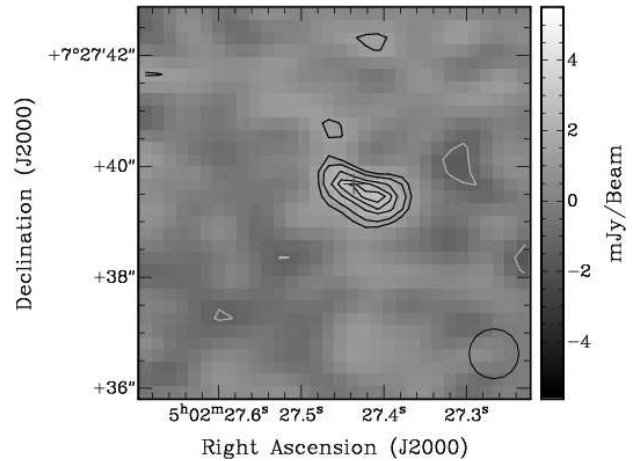} 
\caption{CARMA 1.3-mm continuum map of HD 32297.  The black cross
marks the stellar position, with the full-width representing 
four-times the total rms millimeter positional uncertainty.  Contours
begin at 2$\sigma$ and increase by 1$\sigma$ thereafter ($\sigma=$
0.44 mJy beam$^{-1}, \theta_{\mathrm{beam}}=0.9''$). The source
morphology suggests that HD 32297 may be marginally resolved.  The
measured position angle is consistent with that observed in the NIR
and MIR ($45-57^{\circ}$).  The centroid of the disk emission is
offset from the stellar position at the 4$\sigma$ level.}
\label{map}
\end{figure}

\begin{figure}
\centering
\includegraphics[width=0.18\textwidth,angle=90]{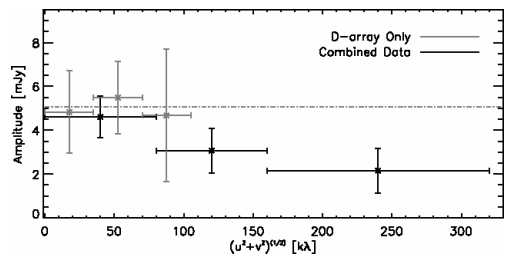}
\caption{Vector-averaged visibility amplitudes as a function of
projected baseline length from phase center.  The combined data
amplitudes decrease with distance from phase center, suggesting the
source is resolved at the 2$\sigma$ level.  The D-array data, on the
other hand, are consistent with a point-source model (grey dashed
line), allowing a robust flux measurement.}
\label{vis}
\end{figure}

Figure \ref{map} shows a 1.3 mm continuum source detected at the
7$\sigma$ level.  The observed peak in the map is $2.8 \pm 0.4$ mJy
beam$^{-1}$.  The probability of detecting an unrelated background
source within 1$''$ of the HD 32297 stellar position is $\ll$1\%,
based on recent source counts at 850 $\mu$m (e.g., Scott et al. 2002).
Thus, we conclude that the detected emission is associated with the HD
32297 debris disk.  The map morphology suggests the source may be
resolved (see below).  However, in the D-array data set alone the
visibility amplitudes are constant with baseline length (Figure
\ref{vis}), suggesting the source is unresolved.  We therefore, adopt
an integrated flux measured using the D-array data set only: $5.1 \pm
1.1$ mJy.

The source centroid in Figure \ref{map} appears offset from the
Hipparcos stellar position (located at phase center).  To quantify
this result, we fit a point source model to the visibility data; the
measured offset is $\Delta r=0.43 \pm 0.08''$ ($\Delta \alpha = -0.30
\pm 0.08''$, $\Delta \delta = -0.31 \pm 0.07''$).  This offset is
significantly larger than that observed for test calibrator, 3C 120,
observed each source cycle ($\Delta r=0.041 \pm 0.001''$, $\Delta
\alpha = -0.010 \pm 0.002''$, $\Delta \delta = 0.040 \pm 0.001''$).
In addition to systematic astrometric uncertainties, we estimate a
statistical positional error for HD 32297 of $\Delta \theta =
(4/\pi)^{1/4} / \sqrt{8 \ln 2} \cdot \theta_\mathrm{beam} / SNR \sim
0.06''$, in agreement with the formal errors derived from fitting the
visibilities.  Combining the systematic and statistical errors, the
total rms positional uncertainty for HD 32297 is $0.09''$.  Thus, the
peak continuum emission is offset from the stellar position at the
4$\sigma$ level.

As previously noted, the Figure \ref{map} map suggests the source is
marginally resolved.  To quantify this effect, in Figure \ref{vis}, we
show the visibility amplitudes for the combined data
set as a function of projected baseline length from phase center.  The
data are binned by baseline length in three circularly symmetric
annular bins, with widths chosen to provide approximately the same
number of visibilities per bin.  The errors in amplitude represent the
standard deviation in the mean of the visibilities in each bin.
Figure \ref{vis} shows that the source amplitudes decline with
baseline length, with a deviation from a point-source model of
approximately 2$\sigma$.  Thus, the source appears to be indeed
resolved, though additional data are needed to confirm this result.
In particular, we note that uncorrected atmospheric phase errors at mm
wavelengths can artificially enlarge targets. Still, the observed
morphology agrees with that observed at other wavelengths.  Fitting an
elliptical Gaussian to the visibilities yields a FWHM of $1.8 \pm
0.6'' \times 0.2 \pm 0.6''$ with a position angle of $55 \pm
10^\circ$, consistent to within 1$\sigma$ of the orientation observed
in the NIR and MIR (P.A.$=45-57^{\circ}$, Schneider et al. 2005, F07,
Moerchen et al. 2007).

\begin{figure}
\centering
\includegraphics[width=0.35\textwidth,angle=0]{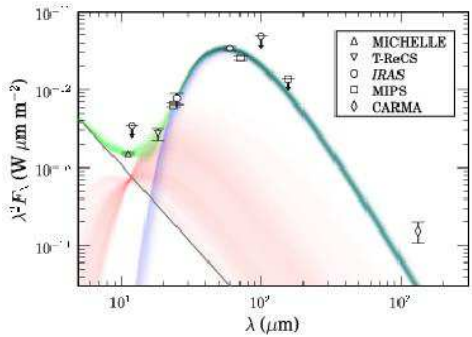}
\caption{Two-population SED fit (see text and F07). The small-grain
  population is plotted in red, the large-grain population in blue,
  the stellar photosphere in grey, and the composite model in green.
  The relative shadings correspond to the 2-d histograms of the
  component SEDs for each model state in the Markov chains.
  Photometry references are as follows: MICHELLE (F07), T-ReCS
  (extracted from data in \citet{Moerchen07}), IRAS \citep{Moor06},
  MIPS (this paper), CARMA (this paper).  The model provides a good
  fit to the mid- and far-infrared data but underestimates the
  observed mm flux, potentially providing evidence for a third grain
  population.}
\label{sed}
\end{figure}

We observe no CO $J=2-1$ line emission in our data.  The 3$\sigma$
limit is 22 mJy beam$^{-1}$ (1.5 K, $\theta_{\mathrm{beam}}=0.70'' \times
0.52''$) in a 42 km s$^{-1}$ channel.  Adopting the same assumptions
as \citet{Dent05}, the corresponding gas mass limit is $M_{gas}
\lesssim 0.3 M_{\mathrm{Jup}}$.

\section{Discussion}

\begin{figure}
\centering
\includegraphics[width=0.23\textwidth,angle=90]{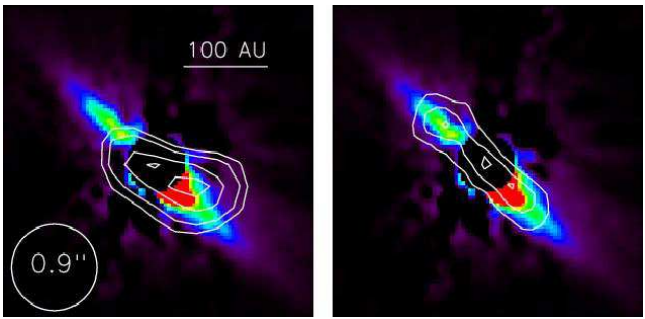}
\caption{{\it Left:} CARMA contours of HD 32297, overlaid on the NIR
scattered-light image from Schneider et al. (2005). {\it Right:}
Photosphere-subtracted MIR contours from F07 overlaid on the same
image. The asymmetry in the CARMA data between the northeast and
southwest lobes suggests the large, mm-sized grains may be trapped in
resonance with an unseen exoplanet.  A similar asymmetry is also
observed in scattered light but not in the MIR.  The contrast between
asymmetry at short and long wavelengths and symmetry at intermediate
wavelengths is a direct prediction of the resonant debris disk models
of Wyatt (2006).}
\label{multiwav}
\end{figure}

F07 attempted to model the observed SED and $N'$-band image of HD
32297 with a single ring of grains of characteristic size, but found
that a second population of grains was needed to adequately fit the
observed SED for $\lambda \gtrsim 25 \mu$m.  Indeed, the contrast
between the observed mm and $N'$-band morphologies (see below and
Figure \ref{multiwav}) suggests that the grains responsible for the
emission at each wavelength constitute separate populations.  To test
whether the population of mm-emitting grains is consistent with the
second, larger grain population proposed by F07 to fit the SED at 25
$\mu$m $\lesssim \lambda \lesssim$ 60 $\mu$m, we revisit the F07
model, adopting their data and fitting method, and incorporating the
Qa-band flux of \citet{Moerchen07} and the new MIPS and CARMA
fluxes. The free parameters in the single population model of F07 are
the disk inclination ($i$), position angle (PA), radii of the inner
and outer edges ($\varpi_0$, $\varpi_1$), surface density power-law
index ($\gamma$), vertical optical depth to absorption at the inner
edge ($\tau_0\equiv\tau_\perp^{\mathrm{abs}}(\varpi_0)$), stellar flux
factor ($\xi$), and effective grain size ($\lambda_\mathrm{sm}$). To
fit the long-wavelength SED ($\lambda \gtrsim 25 \mu$m), we augment
this model with a population of larger grains of effective size
$\lambda_\mathrm{lg}$ and total emitting area $A_\mathrm{lg}$, located
in a narrow ring at the small-grain inner-disk edge, $\varpi_0$.
These grains contribute flux, $F_{\nu,\mathrm{lg}}$, according to:
\begin{eqnarray}
\epsilon_{\nu,\mathrm{lg}} &=&\left\{
\begin{array}{ll}
\lambda_\mathrm{lg}/\lambda & \mbox{if $\lambda>\lambda_\mathrm{lg}$,} \\
1 & \mbox{otherwise},
\end{array}\right. \\
T_\mathrm{lg}(r) &=& 468 \left(\frac{L_*/L_\sun}{\lambda_\mathrm{lg}/1\,\micron}\right)^{1/5} \left(\frac{r}{1\,\mathrm{AU}}\right)^{-2/5} \mathrm{K}, \\ 
F_{\nu,\mathrm{lg}} &=& \left(\frac{A_\mathrm{lg}}{d^2}\right) \epsilon_{\nu,\mathrm{lg}} B_\nu[T_\mathrm{lg}(\varpi_0)].
\end{eqnarray}

Following F07, we ran three Monte Carlo Markov chains with $3\times
10^4$ samples, and simulataneously fit both the large and small grain
populations.  The results of this procedure are listed in Table
\ref{sed_table}, and the corresponding range of allowed dust emission
is plotted in Figure \ref{sed} (red: small grains, blue: large grains,
grey: photosphere, green: composite).  While the two-population model
provides a satisfactory fit to the mid- and far-infrared data, the mm
flux is underestimated at the 4$\sigma$ level, suggesting that the
second grain population proposed by F07 is not responsible for the
majority of the mm flux.  Nevertheless, the new MIPS data lend further
evidence to the F07 suggestion that two populations are needed to fit
the observed SED for $\lambda \lesssim 160 \mu$m.  Therefore, the fit
suggests at least three distinct populations are traced by the current
observations.  However, from the two-population model, only the 1.3 mm
flux appears to trace the putative third population.  Since this
population is described by both a mass/emitting area and a
size/temperature, we lack sufficient data to fully characterize it via
modeling. We note, though, that this population likely traces $\gtrsim
95$\% of the total dust mass.  Assuming a characteristic stellocentric
distance of $\sim$50 AU (\S 3), $L_{*}=5.4L_{\sun}$ (F07), and an
effective grain size of 1.3-mm, the implied mm grain temperature is
$\sim$30 K, suggesting a dust mass of $M_{\mathrm{mm}}\sim M_{\earth}$
(adopting an opacity of 1.7 cm$^2$ g$^{-1}$).  This estimated mass is
among the highest observed for debris disks detected in the (sub)mm
and is two orders of magnitude larger than that implied for the
``large-grain'' population in the SED fit:
$M_{\mathrm{lg}}\sim0.02M_{\earth}$ (using the fitted parameters in
Table \ref{sed_table} and assuming spherical grains with a density of
1 g cm$^{-3}$).  Future far-infrared / sub-mm observations are needed
to confirm the three populations proposed here and better constrain
their properties.

\begin{deluxetable}{lcc}
\tablecaption{\label{sed_table} Best-Fit Model Parameters}
\tablewidth{0pt}
\tablehead{
\colhead{Parameter}  &
\colhead{Best-fit} &
\colhead{Description} }
\startdata
$i$ (deg) & $90  \pm 5$ & disk inclination \\
PA (deg) & $46 \pm 3$ & disk position angle \\
$\varpi_0$ (AU) & $70_{-10}^{+20}$ & inner edge \\
$\varpi_1$ (AU) & $>$ 1200 & outer edge \\
$\log_{10}\tau_0$ & $-2.4_{-0.2}^{+0.3}$ & vertical optical depth at $\varpi_0$ \\
$\gamma$ & $<$ -1.63 & surf density power-law index \\
$\xi / 7.22 \times 10^{-20}$ & $1.03 \pm 0.03$ & $(R_{*}/d)^2$, stellar flux factor \\
$\log_{10}(\lambda_\mathrm{sm}/1\,\micron)$ & $-1.4_{-1.3}^{+0.6}$ & small grain effective size \\
$\log_{10}(A_\mathrm{lg}/1\,\mathrm{cm}^2)$ & $28.9 \pm 0.2$ & large grain emitting area \\
$\log_{10}(\lambda_\mathrm{lg}/1\,\micron)$ & $1.3_{-0.1}^{+0.2}$ & large grain effective size
\enddata
\tablecomments{Confidence intervals are 95\% for marginal
posterior distributions.  Adopted priors are as in F07, except for
$\gamma$, which is constrained to be in the domain [-4,0] due to the
inconsistency of rising surface density with the scattered-light image
and CARMA map; the large grain effective size was constrained by a
log-uniform prior from 1\,nm to 100\,mm and a requirement that
$\lambda_{\mathrm{lg}} > \lambda_{\mathrm{sm}}$.}
\end{deluxetable} 

To address the observed mm morphology, in Figure \ref{multiwav}, we
qualitatively compare the CARMA mm map (contours, left panel) to the
MIR image of F07 (contours, right panel) and the NIR scattered-light
image of \citet{Schneider05} (color, both panels).  The images at
all three wavelengths are consistent with an edge-on disk.  However,
while both the NIR and CARMA data exhibit a brightness asymmetry
between the northeast and southwest lobes inward of $\sim$0.5$''$, the
F07 MIR image is consistent with azimuthal symmetry.  We note that
\citet{Moerchen07} found evidence for asymmetry in their Qa-band data,
though their data were not PSF subtracted, and the asymmetry was in
the opposite sense as observed in the NIR / CARMA data (NE lobe
brighter than SW).

Recently, \citet{Grigorieva07} used predictions from their numerical
model of collisional avalanches to suggest that the observed
scattered-light asymmetry in HD 32297 results from the breakup of a
large planetesimal.  However, while a massive collision can explain
the observed NIR and mm morphology, it is not clear why the MIR image
would not show a similar asymmetry.  An alternative hypothesis is that
the structure results from a planetary-induced resonance.  In this
case, the dust is either trapped in resonance as it drifts inward due
to Poynting-Robertson drag, or it remains locked in resonance after
being generated by parent planetesimals in resonance themselves (e.g.,
Krivov et al. 2007, and references therein).  Interestingly, a recent
study of the latter mechanism by \citet{Wyatt06} directly predicts a
contrast between asymmetry at short and long wavelengths and symmetry
at intermediate wavelengths, as observed in HD 32297.  In this model,
(sub)mm emission is dominated by large grains, which have the same
clumpy resonant distribution as the parent planetesimals.  Small
grains, traced at short wavelengths, exhibit a similar asymmetry, as
they are preferentially born in the high density, resonant structures
before being rapidly expelled from the system.  Lastly,
moderately-sized grains sampled at intermediate wavelengths remain
bound to the star, but have fallen out of resonance due to radiation
pressure and are subsequently scattered into an axisymmetric
morphology.  

This proposed scenario of \citet{Wyatt06} conveniently explains the
qualitative picture for HD 32297 depicted in Figure \ref{multiwav},
yet we caution that this suggestion is highly speculative, and future
rigorous modeling of this system is needed to draw firm conclusions
from the available data.  One potential problem with this hypothesis
is that the SED model predicts that the small ($N'$-band-emitting)
grains have sizes $\lesssim$ 1 $\mu$m (Table \ref{sed_table}), similar
to that expected to produce the NIR scattered-light image (see
discussion in F07).  However, the interpretation of Figure
\ref{multiwav} in terms of the \citet{Wyatt06} models requires that
the NIR-scattering and MIR-emitting grains have different sizes.  This
ambiguity could potentially be resolved through simultaneous modeling
of scattering and emission, incorporating the optical image presented
in \citet{Kalas05}. Most importantly, and independent of speculation,
HD 32297 is currently one of only a few debris disks with resolved
observations in four wavelength regimes (optical, NIR, MIR, mm).
Taken together, these observations can provide a unique testbed for
theories of grain and planetary dynamics.

\acknowledgments H.M. is funded by the GRFP at NSF and the GOPF at UC
Berkeley.  M.\,P.\,F. acknowledges support from the Michelson
Fellowship Program. Support for CARMA construction was derived from
the states of CA, IL, and MD, the Gordon and Betty Moore Foundation,
the Eileen and Kenneth Norris Foundation, the Caltech Associates, and
the National Science Foundation.

{}


\begin{thebibliography}{}
\bibitem[Carpenter et al.(2005)]{Carpenter05} Carpenter, J.~M., et al.\ 2005, \aj, 129, 1049
\bibitem[Corder et al.(2008)]{Corder08} Corder, S.~A., et al.\ 2008, \apjl, submitted
\bibitem[Dent et al.(2005)]{Dent05} Dent, W.~R.~F., et al.\ 2005, \mnras, 359, 663 
\bibitem[Fitzgerald et al.(2007)]{Fitzgerald07} Fitzgerald, M.~P., et al.\ 2007, \apj, 670, 557 
\bibitem[Greaves et al.(1998)]{Greaves98} Greaves, J.~S., et al.\ 1998, \apjl, 5 06, L133
\bibitem[Grigorieva et al.(2007)]{Grigorieva07} Grigorieva, A., et al.\ 2007, \aap, 461, 537 
\bibitem[Holland et al.(1998)]{Holland98} Holland, W.~S., et al.\ 1998, \nat, 392, 788 
\bibitem[Kalas(2005)]{Kalas05} Kalas, P.\ 2005, \apjl, 635, L169 
\bibitem[Krivov et al.(2007)]{Krivov07} Krivov, A.~V. et al.\ 2007, \aap, 462, 199 
\bibitem[Koerner et al.(2001)]{Koerner01} Koerner, D.~W., et al.\ 2001, \apjl, 560, L181 
\bibitem[Moerchen et al.(2007)]{Moerchen07} Moerchen, M.~M., et al.\ 2007, \apjl, 666, L109 
\bibitem[Mo{\'o}r et al.(2006)]{Moor06} Mo{\'o}r, A., et al.\ 2006, \apj, 644, 525 
\bibitem[Moro-Martin et al.(2007)]{MoroMartin07} Moro-Martin, A., et al.\ 2007, ArXiv Astrophysics e-prints, arXiv:astro-ph/0703383 
\bibitem[Perryman et al.(1997)]{Perryman97} Perryman, M.~A.~C., et al.\ 1997, \aap, 323, L49 
\bibitem[Reche et al.(2008)]{Reche08} Reche, R., et al.\ 2008, \aap, 480, 551 
\bibitem[Redfield(2007)]{Redfield07} Redfield, S.\ 2007, \apjl, 656, L97 
\bibitem[Schneider et al.(2005)]{Schneider05} Schneider, G., et al.\ 2005, \apjl, 629, L117 
\bibitem[Scott et al.(2002)]{Scott02} Scott, S.~E., et al.\ 2002, \mnras, 331, 817
\bibitem[Steer et al.(1984)]{Steer84} Steer, D.~G., et al.\ 1984, \aap, 137, 159
\bibitem[Wilner et al.(2002)]{Wilner02} Wilner, D.~J., et al.\ 2002, \apjl, 569, L115 
\bibitem[Wyatt(2006)]{Wyatt06} Wyatt, M.~C.\ 2006, \apj, 639, 1153 
\end{thebibliography}
\end{document}